\documentclass[twocolumn,showpacs,showkeys]{revtex4}
\usepackage{graphicx}

\usepackage{psfrag}
\begin{document}

\title{The Instantaneous Bethe-Salpeter Equation and Its Analog:
the Breit-like Equation}
\author{Chao-Hsi Chang$^{1,2}$ and Jiao-Kai Chen$^{2,3}$}
\address{$^1$
CCAST (World Laboratory), P.O. Box 8730, Beijing 100080,
China\footnote{Not post-mail address.}\\
$^2$ Institute of Theoretical Physics, Chinese Academy of
Sciences, P.O. Box 2735, Beijing 100080, China\\
$^3$ Graduate School of the Chinese Academy of Sciences, Beijing
100039, China}



\begin{abstract}
We take ($\mu^\pm e^\mp$) systems and consider the states with
quantum number $J^P=0^-$ as examples, to explore the different
contents of the instantaneous Bethe-Salpeter (BS) equation and its
analog, Breit equation, by solving them exactly. The results show
that the two equations are not equivalent, although they are
analogous. Furthermore, we point out that the Breit equation
contains extra un-physical solutions, so it should be abandoned if
one wishes to have an accurate description of the bound states for
the instantaneous interacting binding systems.

\end{abstract}


 \maketitle


In the reference \cite{changw1}, we pointed out that the authors
of Refs.\cite{salp,itz} etc replaced the Bethe-Salpeter (BS)
equation which has an instantaneous kernel with an analog of it:
Breit equation \cite{breit,breit1} un-properly. To explore the
different contents i.e. the un-equivalence of the two equations,
in the paper we restrict ourselves to the bound states of the
systems ($\mu^\pm e^\mp$) systems with quantum number $J^P=0^-$
($S$-wave) to solve the two equations exactly and to examine the
obtained solutions accordingly.

BS equation for a fermion-antifermion system has the general
formulation \cite{BS}:
\begin{equation}
(\not\!{p_{1}}-m_{1})\chi_{P}(q)(\not\!{p_{2}}+m_{2})=
i\int\frac{d^{4}k}{(2\pi)^{4}}V(P,q,k)\chi_{P}(k)\;, \label{eq1}
\end{equation}
where $\chi_{P}(q)$ is the BS wave function, $P$ is the total
momentum, $q$ is relative momentum, and $V(P,q,k)$ is the kernel
of the equation, $p_{1}, p_{2}$ are the momenta of the
constituents $1$ and $2$ respectively. The total momentum $P$ and
the relative momentum $q$ are related to the momenta $p_{1},
p_{2}$ as follows:
$p_{1}={\alpha}_{1}P+q, \;\;
{\alpha}_{1}=\frac{m_{1}}{m_{1}+m_{2}},\;\; p_{2}={\alpha}_{2}P-q,
\;\; {\alpha}_{2}=\frac{m_{2}}{m_{1}+m_{2}}.$

If the kernel $V(P,k,q)$ of the BS equation has the behavior:
$V(P,q,k)|_{\vec{P}=0}=V(\vec{q},\vec{k})$ ($\vec{P}=0$ in center
mass frame of the concerned bound state), the BS equation is
called as an `instantaneous BS equation'.

For instance, in Coulomb gauge, the terms corresponding to the
possible transverse-photon exchange between the two components in
BS kernel for ($\mu^\pm e^\mp$) systems are considered as higher
order, so the `lowest order' BS equation kernel for the systems
has the form:
\begin{equation}
V(P,q,k)|_{postronium}=\gamma^0 V_v \gamma^0=-\gamma^0
\frac{4\pi\alpha}{\left(\vec{q}-\vec{k}\right)^{2}}\gamma^0\,.
\label{kernel}
\end{equation}
Namely, there are some physical systems whose bound states are
described by an instantaneous BS equation indeed.

In Refs.\cite{changw1,changw3}, we showed how to solve an
instantaneous BS equation exactly, and also showed the authors of
Ref.\cite{salp,itz} how to mislead the problem to its analog:
Breit equation. For present convenience, let us repeat the main
procedure in Refs.\cite{changw1,changw3}.

Firstly, as done by the authors of Refs.\cite{salp,itz}, the
`instantaneous BS wave function' $\varphi_{P}(\vec{q})$ as
\begin{equation}
\varphi_{P}(\vec{q})\equiv i\int
\frac{dq^0}{2\pi}\chi_{P}(q^0,\vec{q})\,, \label{eqinstw}
\end{equation}
is introduced, then the BS equation Eq.(\ref{eq1}) can be
re-written as
\begin{equation}
\label{eqBSin} \chi_{P}(q^0,\vec{q})=S_f^{(1)}(p^\mu_{1})
\eta(\vec{q})S_f^{(2)}(-p^\mu_{2})\,.
\end{equation}
Here $S_f^{(1)}(p_{1})$ and $S_f^{(2)}(-p_{2})$ are the
propagators of the fermion and anti-fermion respectively and
$$\eta(\vec{q})\equiv\int\frac{d^3\vec{k}}{(2\pi)}
V(\vec{q},\vec{k})\varphi_{P}(\vec{k})\,.$$

The propagators (in C.M.S. i.e. $\vec{P}=0$) can be decomposed as:
\begin{eqnarray}
& -iJS_f^{(i)}(Jp^\mu_{i})=\frac{\Lambda^{+}_{i}(\vec{q})}{Jq^0
+\alpha_{i}M-\omega_i+i\epsilon}+
\frac{\Lambda^{-}_{i}(\vec{q})}{Jq^0+\alpha_{i}M+
\omega_{i}-i\epsilon}\,, \label{eqpropo}
\end{eqnarray}
with $\omega_{i}=\sqrt{m_{i}^{2}+\vec{q}^{2}}\,,\;\;\;\;
\Lambda^{\pm}_{i}(\vec{q})= \frac{1}{2\omega_{i}}\Big[
\gamma^0\omega_{i}\pm J(m_{i} +\vec{\gamma}\cdot\vec{q})\Big]\,,$
where $J=1$ for the fermion ($i=1$) and $J=-1$ for the
anti-fermion ($i=2$). It is easy to check
\begin{eqnarray}
&\Lambda^{\pm}_{i}(\vec{q})+ \Lambda^\mp_{i}(\vec{q})=\gamma^0\,,
\nonumber \\
&\Lambda^\pm_{i}(\vec{q}) \gamma^0
\Lambda^\mp_{i}(\vec{q})=0\,,\;\;\;\;
\Lambda^\pm_{i}(\vec{q})\gamma^0\Lambda^\pm_{i}(\vec{q})
=\Lambda^\pm_{i}(\vec{q})\,.\label{eqpropo1}
\end{eqnarray}
Thus $\Lambda^{\pm}$ can be considered as `energy' projection
operators, and `complete' for the projection. For below
discussions let us introduce the notations
$\varphi^{\pm\pm}_{P}(\vec{q})$ as:
\begin{equation} \label{defini}
\varphi_P^{\pm\pm}(\vec{q})\equiv \Lambda^{\pm}_{1}(\vec{q})
\gamma^0\varphi_{P}(\vec{q})\gamma^0 \Lambda^{\pm}_{2}(\vec{q})\,.
\end{equation}
Because of the completeness of the projection for $\Lambda^{\pm}$,
we have: $$\varphi_{P}(\vec{q})=\varphi^{++}_{P}(\vec{q})
+\varphi^{+-}_{P}(\vec{q})+\varphi^{-+}_{P}(\vec{q})
+\varphi^{--}_{P}(\vec{q})$$ for the BS wave function
$\varphi_{P}(\vec{q})$. If we further carry out a contour
integration for the time-component $q^0$ on both sides of
Eq.(\ref{eqBSin}), then we obtain:
\begin{eqnarray}
\label{eqcont} &
\varphi_{P}(\vec{q})=\frac{\Lambda^{+}_{1}(\vec{q})
\eta_{P}(\vec{q})\Lambda^{+}_{2}(\vec{q})}
{(M-\omega_{1}-\omega_{2})} -\frac{\Lambda^{-}_{1}(\vec{q})
\eta_{P}(\vec{q})\Lambda^{-}_{2}(\vec{q})}
{(M+\omega_{1}+\omega_{2})}\,,
\end{eqnarray}
($M$ is the eigenvalue). To apply the complete set of the
projection operators $\Lambda^\pm_{iP}(\vec{q})$ to
Eq.(\ref{eqcont}), we then obtain the four equations:
\begin{eqnarray}
\label{eqpp} &(M-\omega_{1}-\omega_{2})\varphi^{++}_{P}(\vec{q})=
\Lambda^{+}_{1}(\vec{q})
\eta_{P}(\vec{q})\Lambda^{+}_{2}(\vec{q})\,,
\end{eqnarray}
\begin{eqnarray}
\label{eqmm} &(M+\omega_{1}+\omega_{2})\varphi^{--}_{P}(\vec{q})=
-\Lambda^{-}_{1}(\vec{q})
\eta_{P}(\vec{q})\Lambda^{-}_{2}(\vec{q})\,,
\end{eqnarray}
\begin{eqnarray}
\label{eqpm}
\varphi^{+-}_{P}(\vec{q})=\varphi^{-+}_{P}(\vec{q})=0\,.
\end{eqnarray}
Note that the two equations in Eq.(\ref{eqpm}) do not contain
eigenvalue $M$, so essentially they are constraints for the BS
wave function $\varphi_{P}(\vec{q})$. Because of the completeness
of the projectors $\Lambda^\pm$, only the four equations contained
in Eqs.(\ref{eqpp}, \ref{eqmm}, \ref{eqpm}) are equivalent to the
equation Eq.(\ref{eqcont}).

E.E. Salpeter \cite{salp} and the authors in literature such as
Ref.\cite{itz} would like to connect the above coupled equations
to the Breit equation \cite{breit,breit1}, so they did a
`combination' for Eqs.(\ref{eqpp},\ref{eqmm},\ref{eqpm}) into one
operator equation (in C.M.S. of the bound state, $\vec{P}=0$):
\begin{eqnarray}
&[M-H_1(\vec{q})-H_2(\vec{q})]\varphi(\vec{q}) \nonumber \\
&= \Lambda^{+}_{1}(\vec{q})\gamma^0\eta(\vec{q})\gamma^0
\Lambda^{+}_{2}(\vec{q})-\Lambda^{-}_{1}(\vec{q})
\gamma^0\eta(\vec{q})\gamma^0\Lambda^{-}_{2}(\vec{q})\,,\nonumber\\
&H_1(\vec{q})\equiv m_1\beta+\vec{q}\cdot\vec{\alpha}\,,\;\;\;\;
H_2(\vec{q})\equiv m_2\beta-\vec{q}\cdot\vec{\alpha}\,,
\label{combi}
\end{eqnarray}
with the definitions $\beta=\gamma^0\,,
\vec{\alpha}=\beta\vec{\gamma}$. {\bf In fact, the equation
Eq.(\ref{combi}) is not equivalent to the instantaneous BS
equation i.e. the coupled-equations Eqs.(\ref{eqpp}, \ref{eqmm},
\ref{eqpm}).} When applying the project operator
$\Lambda^{+}_1(\vec{q})\gamma^0\otimes\gamma^0\Lambda^{+}_{2}(\vec{q})$
to Eq.(\ref{combi}) we obtain Eq.(\ref{eqpp}), when applying the
project operator
$\Lambda^{-}_1(\vec{q})\gamma^0\otimes\gamma^0\Lambda^{-}_{2}(\vec{q})$
to Eq.(\ref{combi}) we obtain the Eq.(\ref{eqmm}), whereas when
applying
$\Lambda^{\pm}_1(\vec{q})\gamma^0\otimes\gamma^0\Lambda^{\mp}_{2}(\vec{q})$
to Eq.(\ref{combi}), then we obtain the homogeneous equations:
\begin{eqnarray}
&[M-\omega_1(\vec{q})+\omega_2(\vec{q})]\varphi^{+-}(\vec{q})=0\,,
\nonumber\\
&[M+ \omega_1(\vec{q})
-\omega_2(\vec{q})]\varphi^{-+}(\vec{q})=0\,, \label{eqpm11}
\end{eqnarray}
with $\omega_{1,2}=\sqrt{m^2_{1,2}+\vec{q}^2}$. The two equations
in Eq.(\ref{eqpm11}) contain the eigenvalue $M$, so they do not
play a role as constraints any longer as played by
Eq.(\ref{eqpm}). They are homogeneous equations so we may conclude
that they formally not only have the `trivial solutions' i.e.
Eq.(\ref{eqpm}), but also have `non-trivial solutions'. Note that
when $m_1=m_2$ also means $\omega_1=\omega_2$, so from
Eq.(\ref{eqpm11}) formally one may be sure that only $M=0$ will
correspond to the `non-trivial solutions' of Eq.(\ref{eqpm11}),
whereas, when $M=0$, there is no the frame $\vec{P}=0$ at all, so
Eqs.(\ref{eqpp},\ref{eqmm},\ref{eqpm},\ref{combi},\ref{eqpm11})
written in $\vec{P}=0$ system will not be correct and we have to
re-derive them. To avoid the $m_1=m_2$ case, below we restrict
ourselves to consider the cases $m_1\neq m_2$ only, and it is why
we take ($\mu^\pm e^\mp$) systems as examples below.

It is interesting to explore the differences of the two equations:
the instantaneous BS equation
Eqs.(\ref{eqpp},\ref{eqmm},\ref{eqpm}) and the Breit one
Eq.(\ref{combi}) or say (\ref{eqpp},\ref{eqmm},\ref{eqpm11}),
because it is not only more precisely to prove the un-equivalence
of the two equations than the above argument about the homogeneous
equations Eq.(\ref{eqpm11}), but also is useful for understanding
the two equations: the instantaneous BS equation and the Breit
equation deeply. Specifically, to do it, we restrict ourselves on
the ($\mu^\pm e^\mp$) systems for the bound states with quantum
number $J^P=0^-$ as examples, and solve the coupled equations
Eqs.(\ref{eqpp},\ref{eqmm},\ref{eqpm}) corresponding to the
instantaneous BS equation and
Eqs.(\ref{eqpp},\ref{eqmm},\ref{eqpm11}) corresponding to the
Breit equation respectively without any approximations. In fact,
for the present purpose at the moment, it is enough only to
consider the bound states with quantum number $J^P=0^-$.

As pointed out in Ref.\cite{changw1}, in general, the wave
functions for $J^P=0^{-}$ states in C.M.S. ($\vec{P}=0$) have the
following formulation:
\begin{eqnarray}
\phi&=&\left(\gamma^{0}g_{1}+g_{2}+\hat{q}\hspace{-0.2cm}\slash
g_{3} +\hat{q}\hspace{-0.2cm}\slash \gamma^{0} g_{4}\right)
\gamma^{5}\,,
\end{eqnarray}
where $\hat{q}\hspace{-0.2cm}\slash = \frac{-\vec{q}\cdot
\vec{\gamma}}{|\vec{q}}$. For convenience, let us introduce the
functions $f_i$ ($i=1,2,3,4$) which relate the functions $g_i$ as
the follows:
\begin{eqnarray}
g_{1}&=&f_{1}\,,\nonumber\\
g_{2}&=&f_{2}\,,\nonumber \\
g_{3}&=&\frac{-(m_{1}-m_{2})|\vec{q}|}{m_{1}m_{2}
+\omega_{1}\omega_{2}+\vec{q}^{2}}f_{2}+f_{3}\,,\nonumber \\
g_{4}&=&\frac{(\omega_{1}+\omega_{2})|\vec{q}|}{m_{1}\omega_{2}
+m_{2}\omega_{1}}f_{1}-f_{4}\,.\nonumber
\end{eqnarray}

From Eqs.(\ref{eqpm})), we may straightforward obtain the
requirements $f_3=f_4=0$ and then from Eqs.(\ref{eqpp},\ref{eqmm})
and having the angular integration done, we obtain
\begin{eqnarray}
\displaystyle
Mf_{1}&=&(m_{1}+m_{2})f_{2}-\frac{\alpha_{s}}{\pi}\int
\frac{|\vec{k}|}{|\vec{q}|}d|\vec{k}|\frac{m_{1}+
m_{2}}{2\omega_{1}\omega_{2}(\omega_{1}+\omega_{2})} \nonumber\\
&\cdot&\Big\{(m_{1}m_{2}+\omega_{1}\omega_{2}+\vec{q}^{2})Q_{0}f_{2}\nonumber \\
&+&\displaystyle \frac{(m_{1}-m_{2})^{2}|\vec{q}||\vec{k}|}
{m_{1}m_{2}+E_{1}E_{2}+\vec{k}^{2}}Q_{1}f_{2}\Big\}\,,\label{inBS1}\\
\displaystyle
Mf_{2}&=&\frac{(\omega_{1}+\omega_{2})^{2}}{m_{1}+m_{2}}f_{1}
\nonumber \\
&-&\displaystyle \frac{\alpha_{s}}{\pi}\int
\frac{|\vec{k}|}{|\vec{q}|}d|\vec{k}|\frac{1}{2\omega_{1}\omega_{2}}\cdot
\Big\{(m_{1}\omega_{2}+m_{2}\omega_{1})Q_{0}f_{1}
\nonumber \\
&+&\displaystyle
\frac{(E_{1}+E_{2})(\omega_{1}+\omega_{2})|\vec{q}||\vec{k}|}
{m_{1}E_{2}+m_{2}E_{1}}Q_{1}f_{1}\Big\}\,, \label{inBS2}
\end{eqnarray}
here $E_{i}\equiv
\omega_{i}(|\vec{k}|)=\sqrt{m_{i}^{2}+\vec{k}^{2}}$ and the
angular integrations of the equations have been carried out
already, so $Q_{n}\equiv Q_{n}(\frac{|\vec{q}|^{2}
+|\vec{k}|^{2}}{2|\vec{q}||\vec{k}|})\;\;(n=0,1,\cdots)$, the
$n$th Legendre functions of the second kind appear due to the
precise BS kernel Eq.(\ref{kernel}).

Eqs.(\ref{inBS1}, \ref{inBS2}) are a complete set of coupled
equations about $f_1$ and $f_2$, and exactly equivalent to the
instantaneous BS equation.

While from equations Eqs.(\ref{eqpp},\ref{eqmm},\ref{eqpm11}) i.e.
the Breit equation, with straightforward derivation, we `directly'
obtain the coupled equations:
\begin{eqnarray}
\displaystyle
Mf_{1}&=&(m_{1}+m_{2})f_{2}-\frac{\alpha_{s}}{\pi}\int
\frac{|\vec{k}|}{|\vec{q}|}d|\vec{k}|\frac{m_{1}+
m_{2}}{2\omega_{1}\omega_{2}(\omega_{1}+\omega_{2})} \nonumber\\
&\cdot& \Big\{(m_{1}m_{2}+\omega_{1}\omega_{2}+\vec{q}^{2})Q_{0}f_{2}\nonumber \\
&+&\displaystyle \frac{(m_{1}-m_{2})^{2}|\vec{q}||\vec{k}|}
{m_{1}m_{2}+E_{1}E_{2}+\vec{k}^{2}}Q_{1}f_{2}\nonumber \\
&-&\displaystyle
(m_{1}-m_{2})|\vec{q}|Q_{1}f_{3}\Big\}\,, \label{B-1}\\
\displaystyle
Mf_{2}&=&\frac{(\omega_{1}+\omega_{2})^{2}}{m_{1}+m_{2}}f_{1}
-2|\vec{q}|f_{4}\nonumber \\
&-&\displaystyle \frac{\alpha_{s}}{\pi}\int
\frac{|\vec{k}|}{|\vec{q}|}d|\vec{k}|\frac{1}{2\omega_{1}\omega_{2}}\cdot
\Big\{(m_{1}\omega_{2}+m_{2}\omega_{1})Q_{0}f_{1}
\nonumber \\
&+&\displaystyle
\frac{(E_{1}+E_{2})(\omega_{1}+\omega_{2})|\vec{q}||\vec{k}|}
{m_{1}E_{2}+m_{2}E_{1}}Q_{1}f_{1}\nonumber \\
& -& (\omega_{1}+\omega_{2})|\vec{q}|Q_{1}f_{4}\Big\}\,,
\label{B-2}\\
\displaystyle Mf_{3}&=&\frac{(\omega_{1}-
\omega_{2})^{2}}{m_{1}-m_{2}}f_{4}\,, \label{B-3}\\
Mf_{4}&=&(m_{1}-m_{2})f_{3}\,. \label{B-4}
\end{eqnarray}
Namely Eqs.(\ref{B-1}, \ref{B-2}, \ref{B-3}, \ref{B-4}) are fully
equivalent to Eq.(\ref{combi}) the Breit equation. It is easy to
realize that if one sets $f_3=f_4=0$, the equations
Eqs.(\ref{B-1}, \ref{B-2}, \ref{B-3}, \ref{B-4}) for the Breit
equation will `return' to the equations
Eqs.(\ref{inBS1},\ref{inBS2}) i.e. the instantaneous BS equation.

The normalization condition for the solutions of the equations
reads:
\begin{eqnarray}
&\displaystyle
\int\frac{\vec{q}^{2}d|\vec{q}|}{\left(2\pi\right)^{2}\omega_{1}\omega_{2}}
\Big\{4g_{2}[(m_{2}-m_{1})|\vec{q}|g_{3}\nonumber \\
&+(\omega_{1}+\omega_{2})|\vec{q}|g_{4}
+(m_{1}\omega_{2}+m_{2}\omega_{1})g_{1}]\Big\} =2M.
\end{eqnarray}

\begin{widetext}
\begin{center}
{\bf Table I.} Expansion coefficients for the eigenfunctions (here
$0$ means $0.000\cdots$)
\begin{tabular}{|c|c|c|c|c|c|c|} \hline\hline
 \{nl\} & WF  &$R_{10}$ & $R_{20}$ & $R_{30}$ & $R_{40}$ &$R_{50}$  \\
\hline \cline{1-7}\{10\} &$f_{1}$& 0.707107  & $7.59\cdot 10^{-5}$
& $4.34\cdot 10^{-5}$  & $1.42\cdot 10^{-5}$   & $4.26\cdot 10^{-5}$\\
\cline{2-7} &$f_{2}$& 0.707107  & $7.59\cdot 10^{-5}$
& $4.34\cdot 10^{-5}$  &$1.42\cdot 10^{-5}$   &$4.26\cdot 10^{-5}$   \\
\cline{2-7} &$f_{3}$&0  &0& 0& 0 & 0 \\
\cline{2-7} &$f_{4}$&0&0 &0& 0 & 0       \\
\cline{1-7} \{20\} & $f_{1}$& $-7.98\cdot 10^{-5}$  & 0.707107
& $3.18\cdot 10^{-4}$   & $1.19\cdot 10^{-4}$ & $1.18\cdot 10^{-4}$\\
\cline{2-7} &$f_{2}$&   $-7.98\cdot 10^{-5}$  & 0.707107
& $3.18\cdot 10^{-4} $  & $1.19\cdot 10^{-4}$ & $1.18\cdot 10^{-4}$    \\
\cline{2-7}  &$f_{3}$&0 &0&0&  0       &0        \\
\cline{2-7}  &$f_{4}$&0& 0&0&  0        &0   \\
\cline{1-7}  \{30\} & $f_{1}$& $4.54\cdot 10^{-5}$ &  $3.19\cdot
10^{-4}$
&-0.707106 & $-5.2\cdot 10^{-4}$ & $2.34\cdot 10^{-4}$ \\
\cline{2-7}  &$f_{2}$& $4.54\cdot 10^{-5}$ & $3.19\cdot 10^{-4}$
&-0.707106 &$-5.2\cdot 10^{-4}$ &$2.34\cdot 10^{-4}$  \\
\cline{2-7}  &$f_{3}$&0&0& 0& 0  & 0     \\
\cline{2-7}  &$f_{4}$&0&0&0&  0   &0\\
\cline{1-7}  \{40\} &$f_{1}$& $-1.55\cdot 10^{-5}$& $-1.2\cdot
10^{-4}$ & $-5.21\cdot 10^{-4}$
&0.707106 & $7.66\cdot 10^{-4}$  \\
\cline{2-7}  &$f_{2}$& $-1.55\cdot 10^{-5}$ & $-1.20\cdot 10^{-4}$
& $-5.21\cdot 10^{-4}$  &0.707106 & $7.66\cdot 10^{-4}$ \\
\cline{2-7}   &$f_{3}$ &0 & 0 & 0 &     0      &0         \\
\cline{2-7}  &$f_{4}$&0 &0 &0 &    0      & 0        \\
\cline{1-7}  \{50\} & $f_{1}$&$-4.35\cdot 10^{-5}$ &  $-1.18\cdot
10^{-4}$
&$-2.33\cdot 10^{-4}$  &$-7.67\cdot 10^{-4}$ &0.707106     \\
\cline{2-7}  &$f_{2}$& $-4.35\cdot 10^{-5}$ &  $-1.18\cdot
10^{-4}$
&$-2.33\cdot 10^{-4}$  &$-7.67\cdot 10^{-4}$ &0.707106     \\
\cline{2-7}   &$f_{3}$&0&0 &0&  0&0 \\
\cline{2-7}  &$f_{4}$&0&0 &0 &0  &0   \\
\hline\hline
\end{tabular}
\end{center}
\end{widetext}

Now, let us solve the equations
Eqs.(\ref{B-1},\ref{B-2},\ref{B-3},\ref{B-4}) numerically by
transforming the coupled equations into an eigenvalue problem of a
matrix one, i.e., by expanding $f_{i},\; i=1,2,3,4$ in terms of
the bases of the exact $S$-wave solutions of the Schr\"odinger
equation (in momentum representation)
$R_{nl}(|\vec{k}|)$\cite{onetwo}:
\begin{eqnarray}\label{Schr}
&\displaystyle
R_{nl}(|\vec{k}|)=\sqrt{\frac{2}{\pi}\frac{(n-l-1)!}{(n+l)!}}n^22^{2(l+1)}l!
\nonumber\\
&\displaystyle \cdot
\frac{n^l|\vec{k}|^l}{(n^2|\vec{k}|^2+1)^{l+2}}C^{l+1}_{n-l-1}
\Big(\frac{n^2|\vec{k}|^2-1}{n^2|\vec{k}|^2+1}\Big)
\end{eqnarray}
where $C_N^\nu(x)$ is the Gegenbauer function, defined as the
coefficient of $h^N$ in the expansion of $(1-2hx+h^2)^{-\nu}$ in
powers of $h$. To be practicable, as general cases, the expansion
is truncated according to the request accuracy, so the present
problem becomes an eigenvalue problem of a finite matrix. If we
truncate the expansion up to $j=5$
\begin{eqnarray}
f^{(j)}_i(|\vec{q}|)=\sum_{j=\{nl\}}^{5} C^{(j)}_{i,nl}\cdot
R_{nl}(|\vec{q}|)\,, \label{solution}
\end{eqnarray}
where $j$ denotes the $j$th eigenvalue and eigenfunction, $n$ and
$l$ are the principal and angular quantum numbers: $nl=1S,\, 2S,\,
3S,\, \cdots$ (for $0^-$ states, only $S$-wave states in the
expansion are enough). By making the matrix for the four functions
$f_i,\;(i=1,2,3,4)$, corresponding to a $(4\times 5)\otimes
(4\times 5)$ matrix to be diagonal, we finally obtain the results
(eigenvalues and eigenfunctions accordingly). If ignoring the
negative energy eigenvalue solutions which have similar physics
meanings as those of Dirac equations, we may organize the
solutions into two types: type-A (the so-called `trivial'
solutions $f_3=f_4=0$) and type-B (the so-called `non-trivial'
solutions $f_3\neq 0,\;f_4\neq 0$):

\begin{widetext}
\begin{center}
{\bf Table II.} Expansion coefficients for the eigenfunctions
\begin{tabular}{|c|c|c|c|c|c|c|}
\hline
 \{nl\} & WF  &$R_{10}$ & $R_{20}$ & $R_{30}$ & $R_{40}$ &$R_{50}$  \\
\hline   \{10\}  &$f_{1}$& $2.284\cdot 10^{-3}$  & $1.0679\cdot
10^{-3}$
& $5.770\cdot 10^{-4}$  &$3.615\cdot 10^{-4}$   & $2.407\cdot 10^{-4}$    \\
\cline{2-7} &$f_{2}$& $2.262\cdot 10^{-3}$  & $1.0577\cdot
10^{-3}$
& $5.714\cdot 10^{-4}$    &$3.580\cdot 10^{-4}$   &$2.384\cdot 10^{-4}$   \\
\cline{2-7} &$f_{3}$&-0.615535   &-0.284194
& -0.157829      & -0.101564  & -0.0715402          \\
\cline{2-7}  &$f_{4}$& 0.615535    & 0.284194
&   0.157829    & 0.101564   &0.0715402    \\
\cline{1-7} \{20\} & $f_{1}$&$4.169\cdot 10^{-4}$ &$-3.866\cdot
10^{-4}$
&$-5.112\cdot 10^{-4}$ &$-3.862\cdot 10^{-4}$& $-2.437\cdot 10^{-4}$    \\
\cline{2-7}  &$f_{2}$& $4.129\cdot 10^{-4}$ &$-3.829\cdot 10^{-4}$
&$-5.063\cdot 10^{-4}$  &$-3.825\cdot 10^{-4}$&$-2.414\cdot 10^{-4}$    \\
\cline{2-7}  &$f_{3}$&-0.320488 &0.306022
&0.4.3615       &0.309568  & 0.211896     \\
\cline{2-7}  &$f_{4}$ & 0.320488  & -0.306022
&-0.403615      &-0.309568 &-0.211896       \\
\cline{1-7}  \{30\} & $f_{1}$       &$6.429\cdot 10^{-5}$
&$-3.422\cdot 10^{-4}$
& $-2.655\cdot 10^{-5}$&$2.457\cdot 10^{-4}$ &$2.267\cdot 10^{-4}$  \\
\cline{2-7}  &$f_{2}$ &$6.367\cdot 10^{-5}$ &$-3.389\cdot 10^{-4}$
&$2.630\cdot 10^{-5}$  &$2.434\cdot 10^{-4}$ & $2.245\cdot 10^{-4}$ \\
\cline{2-7}  &$f_{3}$        &-0.119399&0.469465
&$1.961\cdot 10^{-2}$  &-0.364445 & -0.363513 \\
\cline{2-7}  &$f_{4}$        &0.119399&-0.469465
&$-1.961\cdot 10^{-2}$ &0.364445  &0.363513    \\
\cline{1-7}  \{40\} &$f_{1}$&$-2.126\cdot 10^{-5}$ &$1.1331\cdot
10^{-4}$
&$-1.822\cdot 10^{-4}$ &$-3.558\cdot 10^{-5}$   &$1.5455\cdot 10^{-4}$  \\
\cline{2-7}  &$f_{2}$&$-2.106\cdot 10^{-5}$&$1.1222\cdot 10^{-4}$
& $-1.805\cdot 10^{-4}$ &$-3.524\cdot 10^{-5}$ & $1.531\cdot 10^{-4}$ \\
\cline{2-7}  &$f_{3}$&$6.011\cdot 10^{-2}$ &-0.292407
&0.446233 & $7.971\cdot 10^{-2}$ &-0.453219      \\
\cline{2-7}  &$f_{4}$&$-6.011\cdot 10^{-2}$&0.292407
&-0.446233      & $-7.971\cdot 10^{-2}$ &0.453219        \\
\cline{1-7}  \{50\} & $f_{1}$&$1.786\cdot 10^{-7}$ & $2.703\cdot
10^{-5}$
& $-5.647\cdot 10^{-5}$&$9.173\cdot 10^{-5}$ &$-3.973\cdot 10^{-5}$     \\
\cline{2-7}  &$f_{2}$& $1.769\cdot 10^{-7}$ & $2.677\cdot 10^{-5}$
&$-5.593\cdot 10^{-5}$&$9.085\cdot 10^{-5}$&  -3.93438   \\
\cline{2-7}  &$f_{3}$&$2.300\cdot 10^{-2}$  & -0.1403
& 0.335672 & -0.504656  & 0.335311         \\
\cline{2-7}  &$f_{4}$&$-2.300\cdot 10^{-2}$&0.1403
&-0.335672 & 0.504656 &-0.335311      \\
\hline\hline
\end{tabular}
\end{center}
\end{widetext}

{\bf A. The so-called `trivial' solutions}

The `trivial'-eigenvalues (in unit eV) for ($\mu^\pm e^\mp$)
systems are

\noindent $-13.5410866\,,\;\; -3.3863112\,,\;\;-1.5048207\,\\$
$-0.8467912\,,\;\; -0.5420610\,,\;\; \cdots \cdots.$

The eigenfunctions corresponding to the eigenvalues have the
expansion coefficients as those in Table.I.

The over all renormalisation constants for the `trivial' solutions
are $45778.501$, $45778.730$, $45778.772$, $45778.787$
$45778.794$, $\cdots$ accordingly.

In fact, as expected and from Table I one may realize that the
`trivial solutions' of the
Eqs.(\ref{B-1},\ref{B-2},\ref{B-3},\ref{B-4}) have $f_3=f_4=0$
exactly, and indeed they just correspond to the solutions of
Eqs.(\ref{inBS1},\ref{inBS2}).

{\bf B. The so-called `non-trivial' solutions}

The `non-trivial'-eigenvalues (in unit eV) for ($\mu^\pm e^\mp$)
systems are

\noindent $-1.022015012\cdot{10^6}$, $-1.021999529\cdot10^6,\\$
$-1.021998308\cdot10^6,$ $-1.021997962\cdot10^6,\\$
$-1.021997837\cdot 10^6,\;\cdots \cdots.$

\noindent It means that the energy levels have so great binding
energy almost as $2m_e$ (two times of the electron mass). They are
`extra' to the instantaneous BS equation. It seems that in certain
sense, the Breit equation is a `bi-product' of two Dirac
equations, so there are extra solutions, which still `fall' in the
positive energy region for the binding systems, i.e. would
correspond to those negative energy solutions if the lighter
component were free. Thus the gap of the `nontrivial solutions'
from the `trivial ones' can be so great as two times of the
electron mass for the systems ($\mu^\pm e^\mp$). To understand the
fact, we have done several exercises, and find that it is a
general feature for the Breit equation that the `trivial' and
`non-trivial' solutions have so deep a gap as the two times mass
of the lighter component in the bound states.

Since there is no evidence in experiments for the systems
($\mu^\pm e^\mp$), so exactly to say, Breit equation is not
physical. Only when one restrict oneself to consider the week
binding solutions and ignore the (deep) `nontrivial solutions',
the Breit equation is meaningful. In literature, fortunately,
people use Breit equation with a strong implication that only the
very week binding solutions are pursued. Indeed there is no
problem only when applying the Breit equation to the week binding
spectrum studies, whereas if applying it for `complete' energy
spectrum and/or complete set of the wave functions to consider
relativistic correction effects etc, it will cause problem.
However the instantaneous BS equation does not have the problem at
all.

The eigenfunctions corresponding to the eigenvalues have the
expansion coefficients as those in Table.II.

The over all renormalisation constants for the `non-trivial'
solutions are $\\1.240240535\cdot 10^7$, $3.652200818\cdot 10^7$,
$6.728317282\cdot 10^7$, $1.215161253\cdot 10^8$,
$2.757689877\cdot 10^8$, $\cdots $ accordingly.

To test the `stability' of the numerical solutions, we also try to
truncate the expansion up to $j=10$, and obtain the eigenvalues
for `non-trivial' and `trivial' as the follows:

\noindent $-1.022015269\cdot 10^6\,,\;\; -1.021999737\cdot
10^6,\\$ $-1.021998500\cdot 10^6\,,\;\;-1.021998133\cdot 10^6,\\$
$-1.021997979\cdot 10^6\,,\;\;-1.021997900\cdot 10^6,\\$
$-1.021997856\cdot 10^6\,,\;\; -1.021997830\cdot 10^6,\\$
$-1.021997815\cdot 10^6\,,\;\; -1.021997807\cdot 10^6,\\
\cdots \cdots \\$ $-13.5410866\,,\;\;-3.3863112\,,\;\;
-1.5048207,\\$ $-0.8467912\,,\;\;-0.5420610\,,\;\;-0.3764843,\\$
$-0.2764580\,,\;\; -0.2116343\,,\;\; -0.1672546,\\$ $-0.1353994\,,
\;\;\;\;\cdots \cdots.\\ $ From the eigenvalues obtained by a
different truncation, one may see the low-laying numerical
solutions for $j=5$ are quite accurate for our observation. The
table for $j=10$ eigenfunction coefficients is too big to present
here, thus we omit it.

From the example, we have learnt that the precise numerical
solutions of the equations
Eqs.(\ref{B-1},\ref{B-2},\ref{B-3},\ref{B-4}) have confirmed the
arguments on the Eqs.(\ref{eqpm11}) about the homogeneous
equations. Generally, the Breit equation Eq.(\ref{combi}) contains
not only all the solutions of the instantaneous BS equation, but
also the un-physical solutions, the `non-trivial' solutions,
although we have the lessen only for $J^P=0^-$ states of the
($\mu^\pm e^\mp$) systems. Therefore, we can conclude that the
Breit equation Eq.(\ref{combi}) is not equivalent to the
instantaneous BS equation, and the Breit equation Eq.(\ref{combi})
should be abandoned for all kinds of instantaneous interacting
binding systems, if one wishes to apply it to finding exact and
complete solutions and/or uses it for computing the relativistic
correction effects without restriction. In contrary, the
instantaneous BS equation does not have such un-physical solutions
for the instantaneous interacting binding systems at all,
therefore one may take liberties with applying the BS equation to
various problems of the systems.

\vspace{4mm} \noindent {\Large\bf Acknowledgement}: The author
(C.H. Chang) would like to thank Stephen L. Adler for valuable
suggestions and discussions. This work was supported in part by
the National Natural Science Foundation of China.

\end{document}